\documentstyle[12pt]{article}
\author{V.I.Nazaruk\\
Institute for Nuclear Research of RAS, 60th October\\
Anniversary Prospect 7a, 117312 Moscow, Russia\\
(e-mail: nazaruk@inr.msk.su)}
\date{}
\title{Reason of Large Scale Difference of the Results in the $n\bar{n}$
Oscillation Problem}

\begin{document}

\maketitle
\begin{abstract}

We expose a grave drawback hidden in the standard model of $n\bar{n}$
transitions in a medium. Correcting it, one improves a limit on the
free-space $n\bar{n}$ oscillation time $\tau_{n\bar{n}}$ by 31 orders of
magnitude.
\end{abstract}

\vspace{5cm}

INR-0998/May 1995

\newpage

\setcounter{equation}{0}

Recently, the $n\bar{n}$ transitions in a medium beyond the potential model
have been considered[1]. For lower limit $\tau_{min}$ on the free-space
$n\bar{n}$ oscillation time we have got $\tau_{n\bar{n}}>\tau_{min}=4.7\cdot
10^{31}y$, which increases the previous one by 31 orders of magnitude.
Certainly, the value $10^{31}$ provokes distrust. In this letter we explain
in very simple terms the origin of this perceptible (and expected by us)
disagreement.

In the standard approach (labelled bellow as potential model) the $n\bar{n}$
transitions in a medium are described by Schrodinger equations

\begin{equation}
\label{1}(i\partial_t+\nabla^2/2m-U_n)n(x)=\epsilon \bar{n}(x),\;\;\;
(i\partial_t+\nabla^2/2m-U_{\bar{n}})\bar{n}(x)=\epsilon n(x).
\end{equation}
Here $\epsilon=1/\tau_{n\bar{n}}$ is a small parameter[2]; $U_n$ and
$U_{\bar{n}}$ are the self-consistent neutron potential and $\bar{n}$-nucleus
optical potential respectively. For $U_n=const.$ and $U_{\bar{n}}=const.$ in
the lowest order on $\epsilon$ the probability of the process is

\begin{equation}
\label{2}W_{pot}(t)=2ImT_{ii}(t),\;\;\;T_{ii}(t)=i(\epsilon/\delta U)^2[1-
i\delta Ut-\exp (-i\delta Ut)],
\end{equation}
where

\begin{equation}
\label{3}\delta U=U_{\bar{n}}-U_n,\;\;\;U_{\bar{n}}=ReU_{\bar{n}}-i\Gamma /2,
\end{equation}
$\Gamma\sim 100 MeV$ is the annihilation width of $\bar{n}$-nucleus
state. The limit for $\tau_{n\bar{n}}$ is obtained from the inequality
$W_{pot}(T)<1$, where $T\sim 10^{31}y$ is the experimental bound for
the nuclear annihilation lifetime. Priviously, we adduced a physical arguments
that the model (1) is invalid[1,3]. Now we will pinpoint the origin of error
exactly.

$W_{pot}$ describes the $n\bar{n}$-transition, annihilation. Let us consider
the second stage - annihilation decay of $\bar{n}$-medium state. The wave
function of initial state obeys equation

\begin{equation}
\label{4}H_0\Phi =\epsilon_n\Phi,\;\;\;H_0=-\nabla^2/2m+U_n.
\end{equation}
In $\tau =0$ the interaction $\delta U$ is turned on. We have

\begin{equation}
\label{5}\frac{\partial \Psi }{\partial t}=(H_0+\delta U)\Psi ,
\end{equation}
$\Psi(0)=\Phi $. The projection to the initial state is

\begin{equation}
\label{6}<\Phi \mid \Psi (\tau)>=U_{ii}(\tau)=\exp (-i\delta U\tau),
\end{equation}
where $U(\tau)=1+iT^{\bar{n}}(\tau)$ is an evolution operator, $T^{\bar{n}}
(\tau)$ is a transition matrix. The decay probability $W_{\bar{n}}(\tau)$ can
be obtained by two ways.

{\bf 1}. First one gives correct result:

\begin{equation}
\label{7}W_{\bar{n}}(\tau)=1-\mid U_{ii}(\tau)\mid ^2=1-e^{-\Gamma \tau}.
\end{equation}

{\bf 2}. Taking into account that {\em unitarity} condition $UU^+=1$ must be
fulfilled for $U$-matrix of {\em any process} we have

\begin{equation}
\label{8}W_{\bar{n}}(\tau)=2ImT_{ii}^{\bar{n}}(\tau).
\end{equation}
Substituting $T_{ii}^{\bar{n}}=i[1-\exp (-i\delta U\tau )]$ from Eq.(6), one
obtains

\begin{equation}
\label{9}W_{\bar{n}}(\tau)=2\left[1-e^{-\Gamma \tau/2}\cos (\tau Re
\delta U)\right],
\end{equation}
which strongly differs from (7). When $\tau>>1/\Gamma \sim 10^{-24}s$,
Eq.(9) gives

\begin{equation}
\label{10}W_{\bar{n}}=2.
\end{equation}
The probability nonconservation was to be expected because $U_{\bar{n}}$ is
non-Hermitian and unitarity condition is strongly violated: $UU^+=e^{-\Gamma
\tau}\rightarrow 0$. As we will see bellow such is indeed the case in the
calculation of (2). In the slight absorption region  $\Gamma \tau \sim \mid
\delta U\tau\mid<<1$ the probability nonconservation is small: $UU^+=1-\Gamma
 \tau \approx 1$. As a consequence of this Eq.(9) gives result identical to
Eq.(7) one: $W_{\bar{n}}=\Gamma \tau $.

Therefore, the matrix element $ImT_{ii}^{\bar{n}}$ constructed with the help of
non-Hermitian operator $U_{\bar{n}}$ has a sense only when $W_{\bar{n}}=\Gamma
\tau <<1$, or, what is the same

\begin{equation}
\label{11}2ImT_{ii}^{\bar{n}}(\tau)<<1.
\end{equation}
When $\Gamma \tau>1$, expression for $ImT_{ii}^{\bar{n}}$ (i.e.,Eq.
(9)) must be replaced by (7), which is what we made in Ref.[1]. The
alternative is to calculate $ImT_{ii}^{\bar{n}}$ beyond the potential model.
In any case the equality $2ImT_{ii}^{\bar{n}}(\tau)=1-\exp (-\Gamma \tau)$ is
a test for $T_{ii}^{\bar{n}}$ calculation.

Let us return to Eq.(2). For the whole process probability we know full well
that $W_{pot}<<1$. The probability nonconservation is also small $UU^+=1-4
\epsilon ^2t/\Gamma \approx 1$. Thus, as is easy to verify, the both methods
give the same result (2). However, the matrix element $T_{ii}$
itself, and with it the process probability $W_{pot}$ are in error because
they are calculated by means of the erroneous "decay law" (9) in the region
$\Gamma \tau>>1$. In the potential approach this fact is strongly hidden. In
the approach with finite time interval[1] it is obvious.

Really, instead of (2) we have[1]

\begin{equation}
\label{12}W(t)=W^a(t)+W^b(t),\;\;\;\;\;W^a(t)=\epsilon^2t^2\,
\end{equation}
\begin{equation}
\label{13}W^b(t)=-\epsilon^2\int_0^tdt_{\alpha}\int_0^{t_{\alpha}}dt_{\beta}%
W_{\bar{n}}(t_{\alpha}-t_{\beta}),
\end{equation}
where $W^a$,$W^b$  are the contributions of Fig.2a and 2b of Ref.[1]
respectively, $W_{\bar{n}}(\tau)$ is the probability of $\bar{n}$-nucleus
decay in time $\tau=t_{\alpha}-t_{\beta}$. Substituting Eq.(9) in (13), one
obtains

\begin{equation}
\label{14}W(t)=\epsilon^2t^2-\epsilon^2t^2+W_{pot}(t)=W_{pot}(t).
\end{equation}
Consequently, the potential model result is really obtained by means of
"decay law" (9) and in doing so the main contribution gives the range where
$W_{\bar{n}}\approx 2$. (The latter is apparent from (13), where $t=
T=6.5\cdot10^{31}y$[4].) Certainly, this can be understood without going
beyond the potential model framework. Indeed, solving the system (1) by
method of Green functions we get

\begin{equation}
\label{15}T_{ii}(t)=i\epsilon^2\int_0^tdt_{\alpha}\int_0^{t_{\alpha}}dt_
{\beta}[1+iT_{ii}^{\bar{n}}(t_{\alpha}-t_{\beta})].
\end{equation}
Now we have to ask himself: What is used as matrix element of annihilation
decay $T_{ii}^{\bar{n}}$ in (15)? Eq.(10) provides the answer.

The conclusion that potential model is inapplicable to the problem under
study[1] means as follows. Obviously from the point of view of microscopic
theory the potential approach is in principal invalid. Certainly any
characteristic can be parametrized. Eq.(7) is an example of useful
parametrization.  However, instead of it erroneous Eq.(9) is used.
This inevitably arises in solving Eqs.(1), because $2ImT_{ii}$ is represented
through $2ImT_{ii}^{\bar{n}}$, other than $1-\mid U_{ii}(\tau)\mid ^2$.

Our result is obtained by substituting of exponential decay law (7) in (13):

\begin{equation}
\label{16}W(t)=\epsilon^2t^2\left[\frac{1}{2}+\frac{1}{\Gamma t}+\frac{1}%
{\Gamma^2t^2}\left(e^{-\Gamma t}-1\right)\right].\,
\end{equation}
Moreover, in Eq.(13) {\em one can put} $W_{\bar{n}}=const\leq 1$. (We would
like to stress this circumstance.) Then $W\geq \epsilon^2t^2/2$. In this case
{\em only unitarity condition} was employed in deriving of $W(t)$.

It remains to see the reason of enormous quantitative disagreement between the
our and potential model results. The formal answer is that the overall factor
2 in Eq.(9), compared to Eq.(7), leads to the full cancellation of the
$\epsilon^2t^2$ terms in Eq.(14). The strong result sensitivity was to be
expected. Really, the $S$-matrix amplitude $M_s$, corresponding to $n\bar{n}$
transition, annihilation diverges (see Fig.1 and Eq.(5) of Ref.[1]):

\begin{equation}
\label{17}M_s=\epsilon\frac{1}{\epsilon_n-{\bf p}^2_n/2m-U_n}M\sim \frac{1}{0},
\end{equation}
where $M$ is the annihilation amplitude. This is infrared singularities
conditioned by zero momentum transfer in the $\epsilon $-vertex. It is easy to
understand that $M_s\sim 1/0$ for any bound state wave function of neutron
(i.e., for any nuclear model). On the other hand from Eqs.(1) it is clear that
in the potential model the energy is not conserved and becomes complex in the
$\epsilon $-vertex $M_A\rightarrow M_A+\delta U$ ($M_A$ is the nuclear mass).
The corresponding antineutron Green function is

\begin{equation}
\label{18}G=1/(\epsilon_n-{\bf p}^2_n/2m-U_{\bar{n}})=1/\delta U.
\end{equation}
$\delta U=0$ is the peculiar point of $M_s$. So $M_s$ is extremely sensitive
to $\delta U$; the result $W(t)$ is extremely sensitive to $W_{\bar{n}}(\tau)$
obtained by means of $\delta U$. (Usually, the $\delta U$-dependence of $G$ is
masked by $q$: $G^{-1}=(\epsilon_n-q_0)-({\bf p}_n-{\bf q})^2/2m-U_n-\delta
U$. We deal with 2-tail and $q=0$\@.)

The $n\bar{n}$ transition, annihilation (two-step nuclear decay) and particles
motion in the classical fields are the different problems. Describing the
first one by Eqs.(1) we understand that this is an effective procedure. From
formal standpoint in the first and second cases the potentials are complex
and real respectively. (Unfortunately, sometimes the literal analogy between
these problems is drawn[5]\@.) If we try to employ the potential description
(1) as effective one we must invoke an additional reasoning for verification.
Otherwise, we can get (10).

Therefore, the potential approach is inapplicable to the problem under study
neither from microscopic theory standpoint nor from phenomenological
parametrization one, because condition (11) is not fulfilled. More expended
conclusion is as follows. The energy gap $\delta U\neq 0$ is responsible for
dramatic process suppression. It is inevitably exists in the potential
model. However, on the reasons presented above, the potential approach is in
principle invalid. This explains essentially different functional structure
of the results: $W(T)/W_{pot}(T)\sim \Gamma T$. If it is remembered that
$T\sim 10^{31}y$ the quantitative distinction becomes clear as well.

\newpage
%\begin{references}

\end{document}